\documentstyle[12pt]{article}
\begin{document}
\begin{center}
{\large {Study of the "Non-Abelian" Current Algebra of a Non-linear $\sigma$-Model}} \vskip 2cm
Subir Ghosh\\
\vskip .3cm
Physics and Applied Mathematics Unit,\\
Indian Statistical Institute,\\
203 B. T. Road, Calcutta 700108, India\\
\end{center}
\vskip 2cm
{\bf Abstract:}\\
 A particular form of non-linear 
$\sigma$-model, having a global gauge invariance, is studied. The detailed discussion on current algebra structures reveals the non-abelian nature of the invariance, with  {\it{field dependent structure functions}}. Reduction of the field theory to a point particle framework yields a non-linear harmonic oscillator, which is a special case of similar models studied before in \cite{car}. The connection with noncommutative
geometry is also established.

\newpage
{\bf{Introduction:}}\\
\noindent In the present paper we  aim to reveal some interesting properties enjoyed by the current algebra of a particular form of Non-Linear $\sigma$-model. Similar type of models became atractive in the late sixties \cite{del} in the description of mesons, as a non-linear realization of chiral groups, ({\it{e.g.}} $SU(2)\times SU(2)$). Later it appeared that if these models are reduced to their corresponding point particle counterpart, (keeping only time derivatives in the field theory Lagrangian from the operational point of view), they gave rise to non-linear harmonic oscillator models with very interesting behavior \cite{lak}. These oscillators were further studied in all their generality by Carinena et.al. \cite{car}.

The $\sigma $-model that we consider here is, in fact, the massless  limit of the models analysed in \cite{del}. But because of this feature, there are additional global gauge invariances associated with it, leading to interesting current algebra structures. The current algebra in this self-interacting model turns out to be non-abelian, with the structure functions being {\it{field dependent}}. Generic forms of non-abelian current algebra have been studied in \cite{sigma} where it was assumed that the $\sigma$-model exibits a global invariance under some internal symmetry group, represented by a (connected) Lie group $G$ with the corresponding Lie algebra $g$ that obeys $[T^a,T^b]=f^{ab}_{c}T^c$ in some arbitrary basis $(T_a)$, with $f_{ab}^{c}$ being the structure {\it{constants}}. We emphasize that in the present work, we have been able to recover all the results of \cite{sigma} with the field dependent structure function $F^{ab}_{c}(\phi )$ replacing the structure constant $f^{ab}_{c}$. Furthermore, our model reduces to a special case of the non-linear oscillator models studied in \cite{car}, which is expected. We will make some brief observations about it towards the end of the paper.

Works of a similar nature have also appeared in  \cite{ref1,ref2} where particle \cite{ref1} and field theoretic \cite{ref2} models have been studied, which are similar to ours. In \cite{ref1} the particle model is studied from the perspective of a quantum superintegrable model and \cite{ref2} consider the duality aspect of different $\sigma $-models. As we will mention, these models have some non-trivial connection with the noncommutative geometry framework.

  \vskip .5cm
{\bf{The $\sigma$-model:}}\\
We start by considering the Lagrangian,
\begin{equation}
{\cal{L}}=\frac{1}{2}\partial^{\mu}\phi^{a}\partial_{\mu}\phi^{a}+\frac{c}{2}\frac{(\phi^{a}\partial^{\mu}\phi^{a})(\phi^{b}\partial_{\mu}\phi^{b})}{1-c\phi^{a}\phi^{a}}.
\label{20}
\end{equation}
 This is a particular form of $\sigma $-model, in the sense that,
\begin{equation}
{\cal{L}}=\frac{1}{2}G^{ab}(\phi)\partial^{\mu}\phi^{a}\partial_{\mu}\phi^{b}~;~~G^{ab}(\phi)=\delta^{ab}+c\frac{\phi^{a}\phi^{b}}{1-c\phi
^a\phi^{a}}. \label{21}
\end{equation}
Recalling earlier works \cite{del}, our model actually relates to the  choice of Gasiorowicz-Geffen coordinates ({{\it{i.e.}} $\Lambda ^{2}(\varphi^{2} )=c$ with $c$ being a constant in the representation of the meson field $S$ as in  Delbourgo, Salam and Strathdee in \cite{del}),
$$S=\sigma +i\tau .\varphi \Lambda (\varphi^{2}).$$

The variational equation of motion is
\begin{equation}
\partial^{\mu}\partial_{\mu}\phi^{a}+2c{\cal{L}}\phi^{a}=0.
\label{23}
\end{equation}
 For a single field, {\it{i.e.}} $a=1$, exploiting  a mode expansion
for $\phi^{a}(x)=\phi^{a}e^{ik.x}$ in the equation of motion
(\ref{23}), we obtain
\begin{equation}
\frac{k^2}{1-c\phi^{2}}\phi =0, \label{24}
\end{equation}
indicating that the "particles" are massless, $k^2=0$.

Let us now discuss the symmetries of the model. Besides the
obvious spacetime symmetries, the model enjoys a global gauge
invariance, with the symmetry transformation,
\begin{equation}
\delta\phi^{a}=\sqrt{1-c\phi^{2}}\epsilon^{a}, \label{25}
\end{equation}
with $\epsilon^{a}$ being an infinitesimal rigid parameter. The
corresponding conserved Noether current is given by,
\begin{equation}
J^{a}_{\mu}=\sqrt{1-c\phi^{2}}[\partial_{\mu}\phi^{a}+c\frac{(\phi\partial_{\mu}\phi)}{1-c\phi^{2}}\phi^{a}]\equiv
\sqrt{1-c\phi^{2}}G^{ab}\partial_{\mu}\phi^{b}. \label{26}
\end{equation}

However, generalizing this global gauge invariance to a local
invariance is tricky. For a local gauge transformation (\ref{25}),
where $\epsilon^{a}(x)$ is not constant, the Lagrangian transforms
nicely,
\begin{equation}
\delta {\cal{L}}=J^{a}_{\mu}\partial^{\mu}\epsilon^{a}(x).
\label{27}
\end{equation}
However, the current transforms in a covariant way,
$$
\delta
J^{a}_{\mu}=c(\phi^{a}\partial_{\mu}\phi^{b}-\phi^{b}\partial_{\mu}\phi^{a})\epsilon^{b}
=\frac{c}{\sqrt{1-c\phi^{2}}}(\phi^{a}J_{\mu}^{b}-\phi^{b}J_{\mu}^{a})\epsilon^{b}$$
\begin{equation}
=
\frac{c}{\sqrt{1-c\phi^{2}}}\epsilon^{abc}\epsilon^{cde}\epsilon^{b}\phi^{d}J^{e}_{\mu}\equiv
F^{abc}(\phi)\epsilon^{b}J^{c}_{\mu}. \label{28}
\end{equation}
We have identified the structure function to be,
\begin{equation}
F^{abc}=\frac{c}{\sqrt{1-c\phi^{2}}}\epsilon^{abd}\epsilon^{dgc}\phi^{g}=\frac{c}{\sqrt{1-c\phi^{2}}}(\phi^{a}\delta^{bc}-\phi^{b}\delta^{ac}),
\label{43}
\end{equation}
with $F^{abc}=-F^{bac}$.

Hence we can consider an interacting Lagrangian ${\cal{L}}_I$ of
the form,
\begin{equation}
{\cal{L}}_I={\cal{L}} +J^{a}_{\mu}A^{a\mu}, \label{29}
\end{equation}
In order to achieve local "gauge" symmetry, $A^{a}_{\mu}$ will
have to transform as,
$$
\delta
A^{a}_{\mu}=-\partial_{\mu}\epsilon^{a}-\frac{c}{\sqrt{1-c\phi^{2}}}\epsilon^{abc}\epsilon^{cde}\epsilon^{b}\phi^{d}A^{e}_{\mu}=-(\partial_{\mu}\epsilon^{a}+F^{abc}\epsilon^{b}A^{c}_{\mu})$$
\begin{equation}
\equiv -(D_{\mu}\epsilon)^a. \label{30}
\end{equation}
It will be interesting to see if  construction of  a locally gauge invariant kinetic term for
$A^{a}_{\mu}$ is possible, which might require a generalization of the real fields $\phi^{a}$  complex fields.
 \vskip .5cm
{\bf{Current algebra for the $\sigma$-model:}}\\
\noindent The canonical definition for the Energy-Momentum  
tensor is,
\begin{equation}
\theta^{\mu\nu}=\frac{\delta {\cal{L}}}{\delta
(\partial_{\mu}\phi^{a})}\partial^{\nu}\phi^{a}-\eta^{\mu\nu}{\cal{L}}.
\label{22}
\end{equation}
In the present case, this leads to
\begin{equation}
\theta^{\mu\nu}=\partial^{\mu}\phi \partial^{\nu}\phi
+c\frac{(\phi\partial^{\mu}\phi)(\phi\partial^{\nu}\phi)}{1-c\phi
^2}-\eta^{\mu\nu}{\cal{L}}, \label{22a}
\end{equation}
which is conserved, symmetric and traceless for $1+1$-dimensions.

We now revert to a Hamiltonian framework, which is suitable for
studying the current algebra. The conjugate momentum,
\begin{equation}
\pi^{a}=\frac{\delta {\cal{L}}}{\delta \dot
\phi^{a}}~~;~\dot\phi^{a}=\pi^{a}-c(\phi .\pi
)\phi^{a}~~;~\pi^{a}=\dot \phi^{a}+c\frac{(\phi
.\dot\phi)}{1-c\phi^{2}}\phi^{a}, \label{32}
\end{equation}
yields the Hamiltonian and total momentum densities,
\begin{equation}
\theta^{00}\equiv
{\cal{H}}=\frac{1}{2}[\pi^{a}\pi^{a}+\partial_{i}\phi^{a}\partial_{i}\phi^{a}-c(\phi
.\pi )^2+c\frac{(\phi\partial_{i}\phi )(\phi\partial_{i}\phi
)}{1-c\phi ^2}], \label{33}
\end{equation}
\begin{equation}
\theta^{0i}\equiv {\cal{P}}^i=\pi^a \partial ^i\phi^a ,
 \label{34}
\end{equation}
which are the generators of time and spatial translations,
respectively.

The gauge current is also expressed in terms of phase space
variables,
\begin{equation}
J_0^a=\sqrt{1-c\phi^2}\pi^a~;~~J^{a}_{i}=\sqrt{1-c\phi^{2}}(\partial_{i}\phi^{a}+c\frac{(\phi\partial_{i}\phi)}
{1-c\phi^{2}}\phi^{a}).
 \label{35}
\end{equation}
We exploit the equal-time canonical Poisson brackets:
\begin{equation}
\{\phi^{a}(x),\pi^{b}(y)\}=\delta^{ab}\delta
(x-y)~;~~\{\phi^{a}(x),\phi^{b}(y)\}=\{\pi^{a}(x),\pi^{b}(y)\}=0.
\label{35a}
\end{equation}
The internal current algebra is "non-abelian" in nature:
$$
\{J_0^a(x),J_0^b(y)\}=c(\phi^a\pi^b-\phi^b\pi^a)\delta(x-y)=\epsilon^{abc}\epsilon^{cde}
\frac{c}{\sqrt{1-c\phi^2}}\phi^dJ_0^e \delta (x-y)$$
\begin{equation}
=F^{abc}J^{c}_{0}\delta (x-y),
 \label{36}
\end{equation}
$$
\{J_0^a(x),J_i^b(y)\}= ((1-c\phi ^2(x))\delta
^{ab}+c\phi^a(x)\phi^b(x))\partial_i\delta(x-y)+2c(\phi^a\partial_i\phi^b-(\phi
\partial_i\phi )\delta^{ab})\delta (x-y)$$
\begin{equation}
 =J^{ab}(y)\partial_{i}\delta (x-y)+F^{abc}J^{c}_{i}\delta (x-y).
\label{41}
\end{equation}
The current algebra closes by considering $J^{ab}$ as a composite
operator \cite{sigma},
\begin{equation}
J^{ab}=(1-c\phi^{2})\delta^{ab}+c\phi^{a}\phi^{b}, \label{42}
\end{equation}
with the algebra,
\begin{equation}
\{J^{a}_{0}(x),J^{bc}(y)\}=c({\sqrt{1-c\phi^{2}}})(F^{abd}J^{cd}+F^{acd}J^{bd})\delta
(x-y). \label{44}
\end{equation}
Rest of the commutators are trivial since they do not involve the
momenta,
\begin{equation}
\{J^{a}_{i}(x),J^{b}_j(y)\}=\{J^{a}_{i}(x),J^{bc}(y)\}=0.
\label{44a}
\end{equation}
Next we come to the diffeomorphism algebra,
$$
\{{\cal {H}}(x),{\cal {H}}(y)\}=({\cal {P}}_i(x)+ {\cal
{P}}_i(y))\partial_i\delta (x-y),$$
\begin{equation}
\{{\cal {H}}(x),{\cal {P}}_i(y)\}=({\cal {H}}(x)+ {\cal
{H}}(y))\partial_i\delta (x-y)~;~~\{{\cal {P}}_i(x),{\cal
{P}}_j(y)\}=0.
 \label{40}
\end{equation}
Finally we consider the mixed commutators:
$$
\{{\cal{H}}(x),J_0^b(y)\}=J_i^b(x)\partial_i\delta (x-y),$$
\begin{equation}
\{{\cal{H}}(x),J^{a}_{i}(y)\}=J^{a}_{0}(x)\partial_{i}\delta
(x-y)-2c(\partial_{0}J^{a}_{i}-\partial_{i}J^{a}_{0})\delta (x-y).
\label{45}
\end{equation}
In (\ref{45}) we have used the relation,
\begin{equation}
\dot \phi^{a}\equiv \{\phi^{a}(x),H\}=\pi^{a}-c(\phi .\pi
)\phi^{a}. \label{46}
\end{equation}
All the brackets with ${\cal{P}}_i$ are conventional in nature
since ${\cal{P}}_i$ has a canonical structure and so they are not
shown explicitly. Hence the general current algebra structures discussed in \cite{sigma} are exactly reproduced in our model with a field dependent non-abelian structure function.
\vskip .5cm
{\bf{Non-linear harmonic  oscillator and relation with Noncommutative space:}}\\
\noindent
To recover the oscillator model, we formally replace the fields $\phi^{a}(x)$ in (\ref{20}) by 
 $X_i(t)$ variables  and interpret the latter as particle coordinates. This gives us  the
Lagrangian,
\begin{equation}
L=\frac{1}{2}\dot X^2+\frac{c}{2}~\frac{(X.\dot X)^2}{1-cX^2},
\label{1}
\end{equation}
where $X_i$ are spatial coordinates, $X^2=X_iX_i~,~X.\dot X=X_i\dot X_i$, and the model is in
arbitrary space dimension. For convenience, we have taken the mass
to be unity and $c$ is a parameter. As mentioned before, this model is a special case of systems discussed in \cite{car} ($\alpha =0$ in the notation of  \cite{car}).
In fact this model can be
studied as a variable mass problem \cite{levy} as well with
interpreting the total lagrangian (\ref{1}) as the kinetic term.
The
equation of motion,
\begin{equation}
\ddot X_i=-2c [\frac{1}{2}\dot X^2+\frac{c}{2}~\frac{(X.\dot
X)^2}{1-cX^2}]X_i \equiv -(2cL)X_i,\label{1a}
\end{equation}
can be reproduced in the Hamiltonian framework with,
\begin{equation}
P_i\equiv \frac{\partial L}{\partial \dot X_i}=\dot
X_i+c\frac{(X.\dot X)}{1-cX^2}X_i, \label{2}
\end{equation}
\begin{equation}
H=\dot X_iP_i-L=\frac{1}{2}[P^2-c(X.P)^2]. \label{3}
\end{equation}
Exploiting the
canonical Poisson Brackets,
\begin{equation}
\{X_i,P_j\}=\delta _{ij}~,~~\{X_i,X_j\}=\{P_i,P_j\}=0. \label{1c}
\end{equation}
and the definition $\dot A=\{A,H\}$ for any  dynamical
variable $A$, we obtain
\begin{equation}
\ddot X_i=-[2cH]X_i. \label{4}
\end{equation}
Notice that in the present case $L=H$ which justifies the particle to be considered as "free".

To focus on the striking feature of the non-linear oscillator, we
restrict to  one dimensional motion and
consider  bounded and periodic  solution of the form, 
\begin{equation}
X_e=\frac{1}{\sqrt{c}}sin(\sqrt{2cE}t), \label{14b}
\end{equation}
where the subscript $X_e$ is put in to remind us of its exotic
nature. This solution corresponds to the energy $E$,
\begin{equation}
E=\frac{1}{2}[P^2-c(X.P)^2]=\frac{1}{2}A^2\omega ^2~,~~ \omega
=\sqrt{2cE}.\label{7b}
\end{equation}

For comparison, we write down the behavior of a {\it{normal}} harmonic oscillator,
\begin{equation}
H=\frac{1}{2}P^2+\frac{1}{2}cX^2, \label{12}
\end{equation}
that has solution,
\begin{equation}
X=\sqrt{\frac{2E}{c}}sin(\sqrt{c}t). \label{13}
\end{equation}
 Notice that for $X_e$ the {\it{amplitude is fixed}},
depending upon $c$ whereas the {\it{frequency is energy
dependent}}, oscillations becoming more rapid with larger energy.
This is qualitatively  different from the behavior of a normal
oscillator and does not reduce to it in any limit.

In order to see the connection with noncommutative geometry, we now discuss the
conserved quantities. Obviously angular momentum $L_{ij}$ is
conserved:
\begin{equation}
L_{ij}=X_iP_j-X_jP_i~;~~\dot L_{ij}=0. \label{15a}
\end{equation}
But there are other conserved quantities $p_i$ as well:
\begin{equation}
p_i=\sqrt{1-cX^2}P_i~;~~\dot p_i=0. \label{16}
\end{equation}

Now we can forge a connection with a particular form of noncommutative space. Notice that $p_i$ are noncommuting,
\begin{equation}
\{p_i,p_j\}=c(x_ip_j-x_jp_i), \label{16a}
\end{equation}
and together with the identification,
\begin{equation}
x_i\equiv \frac{X_i}{\sqrt{1-cX^2}}, \label{17}
\end{equation}
it is easy to derive,
\begin{equation}
\{x_i,x_j\}=0~;~~\{x_i,p_j\}=\delta_{ij}+cx_ix_j. \label{18}
\end{equation}
Thus, (\ref{16a}) and (\ref{18}) generate a particular form of
non-commutative phase space in $(x_i,p_j)$ that is quite well known
in High Energy Physics \cite{sg,sn}. This is actually a
complimentary form of the Snyder algebra \cite{sn} (see Ghosh in
\cite{sg}). For $c=0$ we recover the normal free particle.

The Hamiltonian (\ref{3}) in $(x_i,p_j)$-spacetime turns out to
be,
\begin{equation}
H=\frac{1}{2}[(1+cx^2)p^2-c(x.p)^2], \label{18b}
\end{equation}
with the subsequent equations of motion in Snyder spacetime,
\begin{equation}
\dot x_i=\{x_i,H\}=(1+cx^2)p_i~,~~  \dot p_i=\{p_i,H\}=0.
\label{18a}
\end{equation}
This might be interpreted as a "free" particle in the sense that there is no {\it{external}} force but it is moving in a constant curvature  Snyder
space. One might interpret the non-canonical term in $\dot x_i$ in (\ref{18a}) as an anomalous velocity term or equivalently consider a  variable mass particle.

 \vskip .5cm
{\bf{Conclusion:}}\\
\noindent 
We have studied a particular  non-linear $\sigma$-model, that corresponds to the massless limit of derivative coupling models studied earlier \cite{del}, in meson phenomenology. Apart from the Poincare invariance, the model enjoys a non-abelian global gauge invariance, with {\it{field dependent structure functions}}.

The field theory studied here is worthwhile for the following
reason. In \cite{car} it was pointed out that the non-linear oscillator model
is integrable \cite{das}, in the sense that it contains conserved
quantities in involution, that are same in number as the number of
degrees of freedom. It will be interesting to see whether similar
conclusions can be drawn for the field theory studied here. From
the analysis done so far in this paper, it is not clear whether
this analogy can be extended to the level of integrability for the
field theory.

In the point particle reduction, our model reduces to a particular form non-linear oscillator, studied earlier in \cite{car}.The connection of the present oscillator model with a specific form of noncommutative space have also been revealed.\\
\vskip .4cm
{\it{Acknowledgement}}: We thank the Referee for the constructive comments.

\end{document}